\documentclass[a4paper]{article}

\usepackage[top=7cm,bottom=2.5cm,headheight=120pt,headsep=15pt,left=6cm,right=1.5cm,marginparwidth=4.2cm,marginparsep=0.5cm]{geometry}

\usepackage{marginnote}
\reversemarginpar  % sets margin notes to the left
\usepackage{lipsum} % Required to insert dummy text
\usepackage{calc}
\usepackage{siunitx}
\usepackage{lineno}
\usepackage[utf8]{inputenc}
\usepackage[T1]{fontenc}
\usepackage[
backend=biber,
natbib=true,
sortcites=true,
defernumbers=true,
style=authoryear,
citestyle=authoryear-comp,
maxnames=99,
maxcitenames=2,
giveninits=true,
terseinits=true,
date=year,
url=false
]{biblatex}
\DeclareNameAlias{default}{family-given}
\renewbibmacro{in:}{%
  \ifentrytype{article}{}{\printtext{\bibstring{in}\intitlepunct}}} % remove 'In:' before journal name
\DeclareFieldFormat[article]{pages}{#1} % remove pp.
\AtEveryBibitem{\ifentrytype{article}{\clearfield{number}}{}} % don't print issue numbers
\DeclareFieldFormat[article, inbook, incollection, inproceedings, misc, thesis, unpublished]{title}{#1} % title without quotes
\usepackage{csquotes}
\RequirePackage[english]{babel} % must be called after biblatex
\DeclareBibliographyCategory{ignore}
\addtocategory{ignore}{recommendation} % adding recommendation to 'ignore' category so that it does not appear in the References
\usepackage{nameref}
\usepackage[pdfborder={0 0 0}]{hyperref}  % sets link border to white
\usepackage{graphbox}  % loads graphicx ppackage with extended options for vertical alignment of figures
\usepackage{microtype}
\DisableLigatures[f]{encoding = *, family = * }
\RequirePackage[default,scale=0.90]{opensans}
\usepackage{xcolor}
\definecolor{darkgray}{HTML}{808080}
\definecolor{mediumgray}{HTML}{6D6E70}
\definecolor{ligthgray}{HTML}{d9d9d9}
\definecolor{pciblue}{HTML}{74adca}
\definecolor{opengreen}{HTML}{77933c}
\usepackage{changepage}
\usepackage[aboveskip=1pt,labelfont=bf,labelsep=period,singlelinecheck=off,justification=centering]{caption}

\usepackage{fancyhdr}  % custom headers/footers
\usepackage{lastpage}  % number of page in the document
\fancyhfoffset[L]{4.5cm}  % offsets header and footer to the left to include margin
\renewcommand{\headrulewidth}{\ifnum\thepage=1 0.5pt \else 0pt \fi} % header ruler only on first page

\newcommand{\DOIrecommendationlink}{\href{https://doi.org/\DOIrecommendation}{https://doi.org/\DOIrecommendation}}

\newcommand{\PCI}{Peer Community In Evolutionary Biology}

\newcommand{\beginingpreprint}{
\vspace*{0.5cm}
\begin{flushleft}
\baselineskip=30pt
\marginpar{
\large\textnormal{\color{pciblue}\\RESEARCH ARTICLE}\\
\vspace*{0.5pt}
\\
\includegraphics[align=c,width=0.5cm]{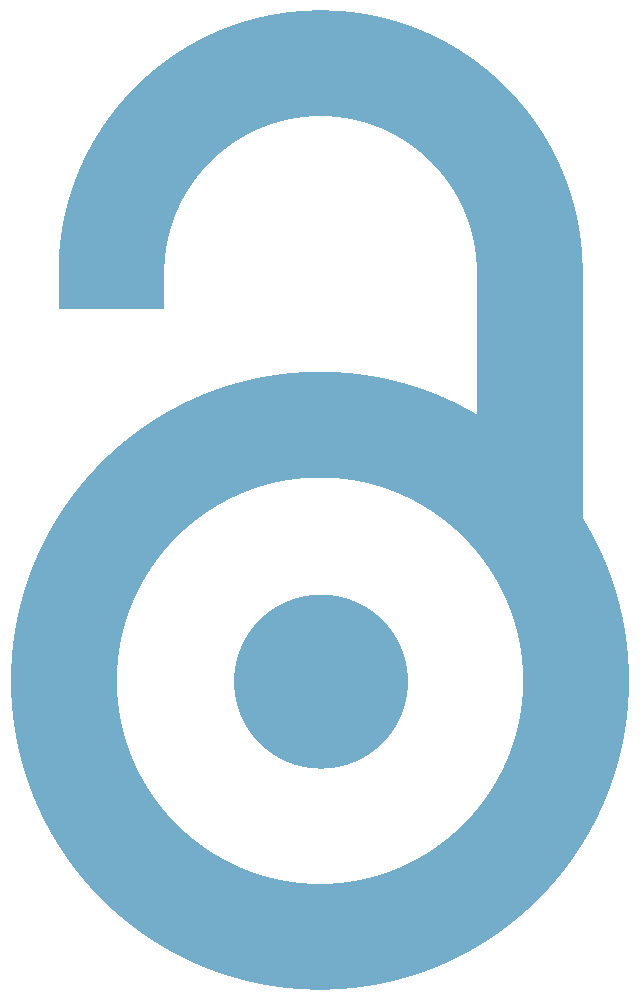} \space \large\textbf{\color{pciblue}Open Access}\\
\\
\includegraphics[align=c,width=0.5cm]{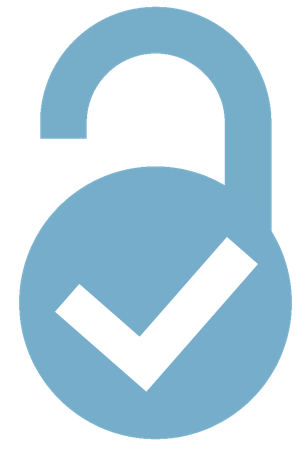} \space \large\textbf{\color{pciblue}Open Peer-Review}\\
\\
\includegraphics[align=c,width=0.5cm]{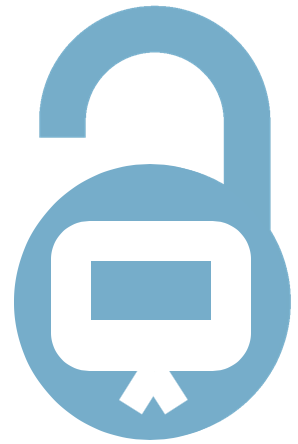} \space \large\textbf{\color{pciblue}Open Code}\\
\\
\\
\\
\\
\raggedright
\scriptsize\textbf{Cite as:}\space
\citeas\\
\vspace*{0.5cm}
\textbf{Posted:} \datepub\\
\vspace*{0.5cm}
\textbf{Recommender:}\\
\recommender\\
\vspace*{0.5cm}
\textbf{Reviewers:}\\
\reviewers\\
\vspace*{0.5cm}
\textbf{Correspondence:}\\
\href{mailto:\email}{\email}\\

}
{\Huge
\fontseries{sb}\selectfont{\preprinttitle}}
\end{flushleft}
\vspace*{0.25cm}
\begin{flushleft}
\Large
\listauthors
\end{flushleft}
\bigskip
{\raggedright
\listinstitutions}
\begin{flushleft}
\fcolorbox{lightgray}{lightgray}{
\parbox{\textwidth - 2\fboxsep}{
\centering\large{\fontseries{sb}\selectfont{This article has been peer-reviewed and recommended by\\
\emph{\PCI} (\DOIrecommendationlink)}}\\
}}
\end{flushleft}
\vspace*{0.5cm}
\fcolorbox{pciblue}{pciblue}{
\parbox{\textwidth - 2\fboxsep}{
\vspace{0.25cm}
\textbf{\large{\textsc{Abstract}}}\\
\preprintabstract\\

\footnotesize{\textbf{\emph{Keywords: }}\preprintkeywords}
\vspace{0.25cm}}
}
}

\newcommand{\preprinttitle}{Gene network robustness as a multivariate character}
\newcommand{\listauthors}{\raggedright 
Arnaud Le~Rouzic\textsuperscript{1}%, \space
}
\newcommand{\listinstitutions}{
\textsuperscript{1} Laboratoire Évolution: Génomes, Comportement, Écologie; Université Paris-Saclay, CNRS, IRD -- Gif-sur-Yvette, France
}
\newcommand{\datepub}{30th March 2022}
\newcommand{\recommender}{Frédéric Guillaume}
\newcommand{\DOIrecommendation}{10.24072/pci.evolbiol.100138}
\newcommand{\reviewers}{Charles Rocabert, Diogo Melo and Charles Mullon}
\newcommand{\citeas}{Le Rouzic A (2022) Gene network robustness as a multivariate character. arXiv:2101.01564, ver. 5 peer-reviewed and recommended by Peer Community in Evolutionary Biology. https://arxiv.org/abs/2101.01564}
\newcommand{\email}{arnaud.le-rouzic@universite-paris-saclay.fr}
\newcommand{\preprintabstract}{
Robustness to genetic or environmental disturbances is often considered as a key property of living systems. Yet, in spite of being discussed since the 1950s, how robustness emerges from the complexity of genetic architectures and how it evolves still remains unclear. In particular, whether or not robustness is independent to various sources of perturbations conditions the range of adaptive scenarios that can be considered. For instance, selection for robustness to heritable mutations is likely to be modest and indirect, and its evolution might result from indirect selection on a pleiotropically-related character (e.g., homeostasis). Here, I propose to treat various robustness measurements as quantitative characters, and study theoretically, by individual-based simulations, their propensity to evolve independently. Based on a simple evolutionary model of a gene regulatory network, I showed that five measurements of the robustness of gene expression to genetic or non-genetic disturbances were substantially correlated. Yet, robustness was mutationally variable in several dimensions, and robustness components could evolve differentially under direct selection pressure. Therefore, the fact that the sensitivity of gene expression to mutations and environmental factors rely on the same gene networks does not preclude distinct evolutionary histories of robustness components.}
\newcommand{\preprintkeywords}{Gene regulatory network; Transcription regulation; Wagner model; Individual-based simulations; Canalization}
\usepackage{bm}
\usepackage{amsmath}

\newcommand{\stability}{{\rho_S}}
\newcommand{\earlyenv}{{\rho_E}}
\newcommand{\lateenv}{{\rho_e}}
\newcommand{\earlymut}{{\rho_M}}
\newcommand{\latemut}{{\rho_m}}

\newcommand{\W}{\bm{\mathrm W}}
\newcommand{\M}{\bm{\mathrm M}}
\newcommand{\Pp}{\bm{\mathrm P}}

\newcommand{\SupMat}{}

\begin{document}
\beginingpreprint

\section*{Introduction}

Robustness is the capacity of living organisms to buffer internal or environmental disturbances. Robustness encompasses, for instance, the ability to maintain physiological equilibria (homeostasis), to ensure developmental stability, or to repair and mitigate DNA damage in both soma and germline. Although robustness is virtually intermingled with the definition of life itself, its underlying mechanisms and its evolutionary origins remain far from being clearly understood \parencite{Ste02,MS09,Wag13,HGK+19}.  \\

Robustness evolves as a consequence of non-linearities in the developmental or physiological mechanisms, i.e.\ changes in the magnitude of the effect of some genetic or environmental factor on the phenotype of interest \parencite{Nij02}. The study of the evolutionary processes leading to robustness roots into the conceptual and empirical work by C.H.\ Waddington and the concept of canalization \parencite{Wad42,Sch49,Wad59,Loi19}. Canalization is a property of complex developmental systems that buffers environmental and genetic variation, and maintains actively the organism in an optimal developmental path. Although the scope and the definition of canalization varies substantially among authors, canalization is generally expected to evolve as an adaptation to "canalizing" selection for an optimal phenotype \parencite{EM98,DD01,Fla05,Kli19}. However, formal population genetic models have questioned the unicity of the canalization process. In particular, robustness to environmental factors appears more likely to evolve as an adaptation than robustness to genetic (mutational) disturbances, on which selection seems to be rather weak and indirect even in optimal theoretical conditions \parencite{WBB97, HHW03,LAH13}.  \\

In this context, the evolution of robustness as a general property of organisms heavily depends on the genetic and physiological integration of the different robustness dimensions \parencite{Far15,FB15}. If the robustness to environmental factors and to genetic mutations share the same physiological bases, the adaptive evolution of environmental canalization can generate a correlated response of genetic canalization; this hypothesis has been referred to as "congruent evolution" \parencite{dHW+03}, and have recieved some empirical support \parencite{Leh10,TES13}. In contrast, if genetic and environmental robustness had independent biological bases, they would be featured by independent evolutionary mechanisms, and possibly independent evolutionary histories.  \\

Although this issue would benefit from a better theoretical framework, modeling the evolution of robustness is not straightforward. The simplest approach relies on modifiers, i.e. genes that can influence the robustness of the organism without affecting the phenotype. However, in the case of genetic robustness, modifier-based models either rely on tricky rescaling or cannot dissociate the phenotype and the robustness to the phenotype \parencite{WBB97, Kaw00, RM13}. In addition, in models where the genotype-phenotype association is arbitrary (such as the NK model, \textcite{KL87}, or the multilinear model, \textcite{HW01}), any correlation between environmental and genetic robustness would be a modeling choice, and not an output of the model. More promising to address the congruent evolution issue are models in which the phenotype is a result of an integrated process mimicking some developmental or physiological mechanism (referred to as \emph{causally cohesive genotype phenotype} models in \cite{RGV08}). In such dynamic models, robustness to various disturbances appear as an emergent property of the model complexity, caused by regulatory feedbacks, that cannot be easily deduced from the model parameters. Although the potential palette of relevant dynamic models is large and could include morphological development models \parencite{MS20}, RNA folding models \parencite{WS99}, or metabolic models \parencite{NBR19}, evolutionary biologists have often considered gene regulatory network models as a good compromise between complexity and numerical tractability for studying the evolution of canalization and robustness \parencite{Kau69,Wag94,SBB00,LP12}. \\

Such theoretical gene networks have been shown to display enough non-linearity, leading to epistasis and pleiotropy, to evolve enhanced or reduced sensitivity to environmental \parencite{Mas04,EMW11,EMW11b} and genetic \parencite{Wag96,BS03,DW08,ALS+06,RL16} perturbations. Interesting observations suggest that environmental or genetic canalization could be correlated to other robustness properties in such models. For instance, \textcite{CMW07,Kan07} noticed that robustness to mutations and robustness to noise was correlated in gene networks --- a similar result was obtained earlier for RNA-folding structures \parencite{Fon02}. Furthermore, it has been shown that network stability, the propensity of the network to maintain stable (non-cyclic) gene expressions, was correlated to robustness, as selection on stability alone could drive an indirect response of genetic \parencite{SB02} and environmental \parencite{Mas04,NK20} canalization. In contrast, \textcite{ORL18} showed that networks selected to maintain (but not converge to) an equilibrium became both environmentally sensitive and genetically canalized, suggesting that environmental and genetic robustness could be theoretically decoupled. However, no systematic quantitative description of the pleiotropic pattern underlying different robustness components has ever been attempted.  \\

Here, I aim at extending the study of canalization in theoretical gene networks to address the multidimensional nature of robustness, by estimating the evolutionary independence of various robustness components. Four robustness-related measurements were considered, two of them corresponding to environmental robustness (early vs.\ late disturbances), two corresponding to genetic robustness (early --- inherited --- or late --- acquired --- mutations). Gene expression instability was also included in the set of robustness-related traits, as it is related to the intrinsic stability of the expression phenotype. The first part of this study focuses on the multidimensional patterns of robustness in small and random networks, and the second part on the evolutionary consequences of the pleiotropic nature of robustness, based on individual-based simulations.  \\

\section*{Model and Methods}

\subsection*{Gene regulatory network}

The network model belongs to the family of gene regulatory network models sometimes referred to as "Wagner model" (after \textcite{Wag94,Wag96}; see \textcite{FP15} for a historical record). Two variants of the model were proposed in \textcite{Wag94}; the second one, involving discrete gene expressions scaled between $-1$ and $1$, has often been reused in the literature \parencite{Wag96,SB02,CMW07}. The model described below is closer to the first model by \textcite{Wag94}, featuring a continuous gene expression $\Pp$ between 0 and 1, and a constitutive expression level $0 < a < 1$ that can be lower than the mid-expression point.  \\

More specifically, the structure of a $n$-gene network is encoded as a $n\times n$ matrix $\W$, while the state of the network is stored into a vector of size $n$, $\Pp$. In this setting, $W_{ij}$ encodes the influence of gene $j$ on the expression of gene $i$, $W_{ij} < 0$ represents a negative interaction (inhibition), $W_{ij} > 0$ a positive interaction (activation), and $W_{ij} = 0$ denotes the absence of regulatory interaction. $P_i$ is the expression of gene $i$, ranging between 0 (no expression) and 1 (maximum expression).  \\

The properties of these gene networks are explored in a discrete dynamic system:
\begin{equation}
 \Pp_{t+1} = F(\W \Pp_t),
\end{equation}
\noindent where the function $F$ is a vectorized version of a sigmoid scaling function: $F(x_1, x_2, \dots, x_n) = [f(x_1), f(x_2), \dots, f(x_n)]$;
\begin{equation} \label{eq:fx}
f(x) = \frac{1}{1+ \lambda_a e ^{- \mu_a x}}, 
\end{equation}
\noindent with $\lambda_a = (1-a)/a$ and $\mu_a = 1/a(1-a)$ \parencite{GCL+18}. The function $f$ is scaled such that $f(0) = a$ and $df/dx|_{x=0}=1$; the parameter $a$ thus stands for the constitutive gene expression (the expression of a gene in absence of regulators), and this function defines the scale of the matrix $\W$: $W_{ij} = \delta$ ($\delta \ll 1$) means that the expression of gene $i$ at the next time step will tend to $P_{i,t+1} = a + \delta$ if $i$ is regulated by a single, fully expressed transcription factor $j$ ($P_{j,t} = 1$). This setting, extensively described in \textcite{RL16}, differs mathematically from the constitutive expression model in \textcite{Wag94} that shifts the sigmoid as $\Pp_{t+1} = F(\W \Pp_t + a)$. \\

Gene networks dynamics start from an initial expression $\Pp_0$, and gene expression was updated for $T$ time steps. By default, $\Pp_0 = (a, a, ..., a)$, since this step immediately follows a virtual initial state with no expression. The expression phenotype corresponding to a gene network was determined by averaging gene expressions during the last $\tau$ time steps for each gene $i$: $P_i^* = (1/\tau)\sum_{t=T-\tau}^T P_{it}$.  \\

\subsection*{Robustness indicators}

Five robustness indicators were calculated, corresponding to five different aspects of genetic or environmental robustness in a gene network: robustness to early ($\earlyenv$) and late ($\lateenv$) environmental disturbance, and robustness to early ($\earlymut$) and late ($\latemut$) genetic disturbance, and network stability $\stability$. All indicators were expressed on a scale homogeneous to log variances in gene expressions; the mode of calculation is summarized in Table~\ref{tab:indicators}, robustness is maximal when the index $\rho$ is small.  \\

The robustness to early environmental disturbance $\earlyenv$ measures the capacity of a network to reach a consistent final state starting from different initial gene expressions. In practice, $R$ replicates of the network dynamics were run, in which the initial gene expressions ($\Pp_0$) were drawn into Gaussian ($\mu = a, \sigma = \sigma_E$) distributions (expression values $<0$ and $>1$ were set to $0$ and $1$, respectively). The environmental robustness $\earlyenv_i$ for each gene $i$ was measured as the log variance in the final gene expression across these replicates.  \\

The robustness to late environmental disturbance $\lateenv$ measures the capacity of a network to recover its equilibrium state after having being disturbed. Gene expressions after $T$ time steps were disturbed by adding a random Gaussian noise of standard deviation $\sigma_e$ to each gene of the network, and $\lateenv_i$ was computed for each gene $i$ as the log variance in gene expression at time step $T+1$ over $R$ replicates.  \\

The robustness to early mutations $\earlymut$ measures the system robustness to inherited genetic mutations (modifications of the $\W$ matrix). A random non-zero element of the $\W$ matrix was shifted by a random Gaussian number of standard deviation $\sigma_M$, and its consequences on the mean expression of all network genes was recorded. The procedure was replicated $R$ times, and the robustness score $\earlymut_i$ for each gene $i$ was calculated as the log variance of gene expression across $R$ replicates.  \\

The robustness to late mutations $\latemut$ measured the effect of mutations in the gene network $\W$ after having reached the final state. In practice, the $\W$ matrix was mutated in the same way as for $\earlymut$ with a standard deviation $\sigma_m$, but its consequences on gene expression were calculated for only one time step, starting from the last state of the network. The robustness score was calculated as for other indicators (log variance over $R$ replicates). \\

Finally, dynamic systems based on the Wagner model often tend to generate limit cycles and never converge to a stable equilibrium. Network stability $\stability$ quantifies the capacity for a specific network to lead to stable gene expressions.  For consistency with other indicators, this instability was measured as the log squared difference between the average expression during the last $\tau$ time steps, and an extra time step.  \\

\begin{table}
%~ \captionsetup{justification=centering}
\begin{adjustwidth}{-1in}{0in}
\begin{tabular}{lp{3.5cm}ll}
Indicator & Robustness component & Computation & Disturbance std.\ dev. \\ \hline
$\earlyenv$ & \raggedright  Early noise in gene \mbox{expression} & $\earlyenv_i = \log [ \frac{1}{R-1} \sum_{r=1}^R (P_{i,r}^* - \overline{P_{i}^*})^2 ]$ & $\sigma_E=0.1$\\
$\lateenv$  & \raggedright  Late noise in gene \mbox{expression} & $\lateenv_i = \log [ \frac{1}{R-1} \sum_{r=1}^R (P_{i,T+1,r} - \overline{P_{i,T+1}})^2]$ & $\sigma_e=0.1$\\
$\earlymut$ & \raggedright  Early (inherited) \mbox{mutations} & $\earlymut_i = \log[ \frac{1}{R-1} \sum_{r=1}^R (P_{i,r}^* - \overline{P_{i}^*})^2]$ & $\sigma_M = 0.1$\\
$\latemut$ & \raggedright  Late (aquired) \mbox{mutations} & $\latemut_i = \log [ \frac{1}{R-1} \sum_{r=1}^R (P_{i,T+1,r} - \overline{P_{i,T+1}})^2]$ & $\sigma_m = 0.1$ \\
$\stability$ & \raggedright Expression stability & $\stability_i = \log[( P_i^* - P_{T+1})^2]$ & \\
\end{tabular}
\caption{\label{tab:indicators} Summarized calculation of all five robustness indicators. Index $i$ stands for the gene ($1 \leq i \leq n$), and $r$ for the replicate ($1 \leq r \leq R$), since all indicators except $\stability$ were estimated by a resampling procedure. $P_i^*$ stands for the equilibrium gene expression of gene $i$ (mean expression from the last $\tau$ time steps), and $\overline{P_{i}^*} = (1/R)\sum_{r=1}^R P_{i,r}^*$ represents the mean over replicates. Noise in gene expression was simulated by adding a random Gaussian deviation to the initial state $\Pp_0$ of the network (for $\earlyenv$) or to the last state $\Pp_T$ of the network (for $\lateenv$). Mutations were simulated by adding a random deviation to a random interaction in the network $\bm{\mathrm W}$, either before starting the network dynamics ($\earlymut$) or after the last time step ($\latemut$). All robustness indicators are homogeneous to a log variance in gene expression; robustness increases when the indicator gets smaller, and sensitivity increases when the indicator increases. The last column indicates the standard deviation of the corresponding Gaussian disturbance. }
\end{adjustwidth}
\end{table}

All these scores were calculated for every gene $i$ of a given network, and then averaged over all genes in order to get a series of summary network descriptors. The magnitude of the score itself is arbitrary, as it depends on the size of the disturbance. However, indicators happen to increase approximately linearly with the size of the disturbance (\SupMat \ref{supp:sensitmag}), the results were thus largely unaffected by a change in the variance of mutational effects and environmental noise.  \\

\subsection*{Random networks}

Random networks were generated as $n\times n$ $\bm{\mathrm W}$ matrices filled with independent identically-distributed random numbers drawn into a Gaussian (by default: $\mu_0 = 0, \sigma_0 = 1$) distribution.  A density parameter $1/n \leq d \leq 1$ could be specified, corresponding to the frequency of non-zero slots in the $\W$ matrix. Zeros were placed randomly, with the constraint that all genes should be regulated by at least another one.  \\

\subsection*{Exhaustive exploration of two-gene networks}

The main interest of gene-network models is the complexity and the richness of the underlying genotype-phenotype relationship. As a side effect, such models are in general difficult to handle mathematically (\cite{CTH11,LP12}). Excluding the one-gene self-regulating case (which already has non-trivial mathematical properties, \textcite{GCL+18}), the simplest network (2-by-2 matrix) has four genetic parameters, which makes the exploration of the parameter set tedious. Here, the number of dimensions was restricted by considering the set of networks that lead to a predefined arbitrary equilibrium, $\bm \Pp^{\theta}_\infty = (P^{\theta}_1, P^{\theta}_2)$. As $F(\W \Pp^{\theta}_\infty) = \Pp^{\theta}_\infty$, the $\W$ matrix can be reduced to two independent parameters, $W_{11}$ and $W_{21}$:
\begin{equation}
    \W = F \left [\begin{pmatrix} W_{11} & A \\ W_{21} & B \end{pmatrix}  \begin{pmatrix} P^{\theta}_1 \\ P^{\theta}_2 \end{pmatrix} \right] = \begin{pmatrix}P^{\theta}_1 \\ P^{\theta}_2 \end{pmatrix},
\end{equation}
\noindent with
\begin{equation}
	\label{eq:predef}
    \begin{split}
        A = \frac{1}{P^{\theta }_2} [f^{-1}(P^{\theta }_1)-W_{11}P^{\theta }_1], \\
        B = \frac{1}{P^{\theta }_2} [f^{-1}(P^{\theta }_2) - W_{21} P^{\theta }_1],
    \end{split}
\end{equation}
\noindent $f^{-1}(y) = -\frac{1}{\mu_a} \log \left( \frac{1-y}{\lambda_a y} \right)$ being the inverse of $f(x)$ (equation~\ref{eq:fx}). This equation can be extended to any network size, provided that a single element $W_{ij}$ is unknown for each line $i$ of the matrix:

\begin{equation}\label{eq:predefN}
	W_{ij} = \frac{1}{P^{\theta}_j} [f^{-1}(P^{\theta}_i) - \sum_{j^\prime \neq j} W_{ij^\prime} P^{\theta}_{j^\prime} ].
\end{equation}

Among the $n^2$ elements of a $n$-gene network, there are thus $n(n-1)$ neutral dimensions that can be explored without modifying equilibrium gene expressions. Large gene networks are thus characterized by a proportionally larger neutral space. \\

The $\W$ matrix achieving the desired $\Pp^{\theta \ast}_\infty$ equilibrium from a specific pair $W_{11}, W_{21}$ always exists (and is unique), but the stability of the equilibrium is not guaranteed. Networks which final gene expression $\Pp^\ast = (P^\ast_1,P^\ast_2)$ differed substantially from the target (in practice, when $|P^\ast_1 - P^{\theta}_1| + |P^\ast_2 - P^{\theta}_2| > 0.15$) were excluded from the analysis. Such discrepancies correspond to either unstable equilibria (in which case gene expressions were driven away from the equilibrium) or extreme oscillatory behaviors (large oscillations may hit expression limits 0 or 1, which drives the average expression away from the target equilibrium).  \\

\subsection*{Evolutionary simulations}

The evolution of gene networks under various evolutionary constraints was studied by individual-based simulations. Each individual was featured by its genotype (a $n \times n$ $\W$ matrix, by default $n=6$ to limit the computational burden), its expression phenotype $\Pp^*$, and the five robustness scores $\stability$, $\earlyenv$, $\lateenv$, $\earlymut$, and $\latemut$. Individuals were haploid and reproduced clonally. Mutations consisted in adding a random Gaussian deviate of variance $\sigma_\nu^2$ to a random regulatory interaction of the $\W$ matrix, with a rate $\nu$ per individual and per generation. Mutational parameters $\nu$ and $\sigma_\nu$ were kept reasonably low to limit the strength of indirect selection for genetic robustness \parencite{WBB97,RL16}. Generations were non-overlapping, and population size $N$ was constant. A generation consists in sampling $N$ new individuals among the $N$ parents, with a probability proportional to the individual fitness. Fitness was computed assuming stabilizing selection around a target (optimal) expression level for $n^\prime \leq n$ genes of the network (by default $n^\prime=3$), as $w = \exp(- \sum_{i=1}^{n^\prime} s_i (P_i^* - \theta_i)^2 )$, where $s_i$ was the strength of stabilizing selection on gene $i$ ($s_i = 0$ standing for no selection), and $\theta_i$ was the optimal expression phenotype. The $\theta_i$ were drawn in a uniform (0,1) distribution at the beginning of each replicated simulation, and the initial gene network was empty ($W_{ij} = 0$) except for one random element per line, which was initialized to match the optimal expression using equation~(\ref{eq:predefN}).  \\

The evolution of robustness components was tracked by estimating $\stability$, $\earlyenv$, $\lateenv$, $\earlymut$, and $\latemut$ at regular time points. Components were estimated for each individual, and averaged out over the population. The response to direct or indirect selection was computed as the average change from generation 0; the multivariate response was stored as a 5-dimension vector $\bm{\mathrm R}$. Simulation runs were replicated 100 times and the results were averaged out, default parameter values are provided in Table~\ref{tab:defpar}.  \\

\begin{table}
%~ \captionsetup{justification=centering}
\begin{tabular}{lll}
Parameter & Symbol & Value \\ \hline
Population size & $N$ & $1000$ \\
Gene network size & $n$ & $6$ \\
Constitutive expression & $a$ & $0.2$ \\
Network time steps & $T$ & $16$ \\
Network measurement steps & $\tau$ & $4$ \\
Network density & $d$ & $1.0$ \\
Simulation replicates & & $100$ \\
Mutation rate per individual & $\nu$ & $0.01$ \\
Size of mutational effects & $\sigma_\nu$ & $0.1$ \\
Number of selected genes & $n^\prime$ & $3$ \\
Stabilizing selection coefficient & $s$ & $10$ \\
Directional selection coefficient & $\beta$ & $0$ \\
Number of robustness tests & $R$ & $100$ \\
Size of early environmental noise & $\sigma_E$ & $0.1$ \\
Size of late environmental noise & $\sigma_e$ & $0.1$ \\
Size of early genetic mutations & $\sigma_M$ & $0.1$ \\
Size of late genetic mutations & $\sigma_m$ & $0.1$ \\
\end{tabular}
\caption{\label{tab:defpar} Default parameter values in the evolutionary simulations.}
\end{table}

Directional selection on robustness indicators was also performed in some simulations, consisting in multiplying individual fitness by $\exp ( \sum_{x \in (S, E, e, M, m)} \beta_x \rho _x )$, where $\beta_x$ was the strength of directional (positive or negative) selection on robustness index $x$ (in practice, $\beta_x= \pm 0.01$). The vector $\bm\beta$ is thus proportional to the multivariate selection gradient on robustness components. There was no correlated selection (the fitness function is the product of independent marginal functions applied on gene expressions and robustness components).  \\

Estimating genetic covariance matrices $\bm{\mathrm G}$ was computationally intractable in simulations (it would require a heavy resampling procedure in each individual), mutational covariances $\M$ from the average genotype in the population ($\overline{\W}$) were used instead to derive multivariate evolutionary predictions. Mutational covariance matrices $\M = \nu \bm{\mathrm C}/5$ were estimated from covariances $\bm{\mathrm C}$ in gene expressions and robustness coefficients among 100 gene networks differing from $\overline{\W}$ by 5 mutations (drawn from the same algorithm as during the simulations). In order to control for the influence of stabilizing selection on gene expression on the evolution of robustness, conditional mutational matrices (equivalent to conditional evolvabilities of $ \bm{\mathrm G}$ matrices in \cite{HH08}) were computed as $\M_{c(y|x)} =\M_y - \M_{yx} \M_x^{-1} \M_{xy}$, where $y$ indicate the $n_y$ unconstrained traits and $x$ the $n_x$ constrained traits (i.e.\ the $n^\prime = 3$ genes under stabilizing selection). $\M_{c(y|x)}$ was thus a $n_y \times n_y$ matrix measuring how the unconstrained traits can mutate while traits $x$ remain constant. Predicted mutational evolvabilities in the direction of selection $\bm \beta$ were calculated as $e_\mathrm{pred} = \bm\beta^\top \M_c \bm\beta / |\bm\beta|^2$ \parencite{HH08}, and realized (observed) evolvabilities were obtained by projecting the multivariate response to selection $\bm{\mathrm R}$ on the direction of $\bm\beta$: $e_\mathrm{obs} = \bm{\mathrm R} \bm\beta / |\bm\beta|$. Contrary to the genetic covariances $\bm{\mathrm G}$, mutational covariances $\M$ cannot be used directly to compute quantitative evolutionary predictions, as the relationship between $\M$ and $\bm{\mathrm G}$ depends on the mutation-selection-drift equilibrium, which is notoriously difficult to handle mathematically \parencite{BL94}. The following analyses thus focus on whether mutational evolvabilities are proportional to the selection responses, assuming that $\bm{\mathrm G}$ are proportional to $\M$.   \\

Simulations and data analysis were coded in R \parencite{R20}, except for the core gene network dynamics that was coded in C++ and embedded in the R code with the Rcpp package \parencite{EF11}. Scripts to reproduce simulations and figures are available online (\url{https://doi.org/10.5281/zenodo.6393075}), \textcite{GitHubRepos}.  \\

\section*{Results}

\subsection*{Random networks}

Random interaction matrices are regularly used in the literature to study the general properties of gene networks (e.g.\ \cite{CTH11,PBF12}). As such, random networks are not expected to reflect the properties of biologically-realistic genetic architectures, as biological networks are far from random. However, such an approach helps developing a general intuition about the properties of the underlying model.  \\

\begin{figure}[t]
\begin{adjustwidth}{-1.5in}{0in}
%~ \captionsetup{justification=centering}
\includegraphics{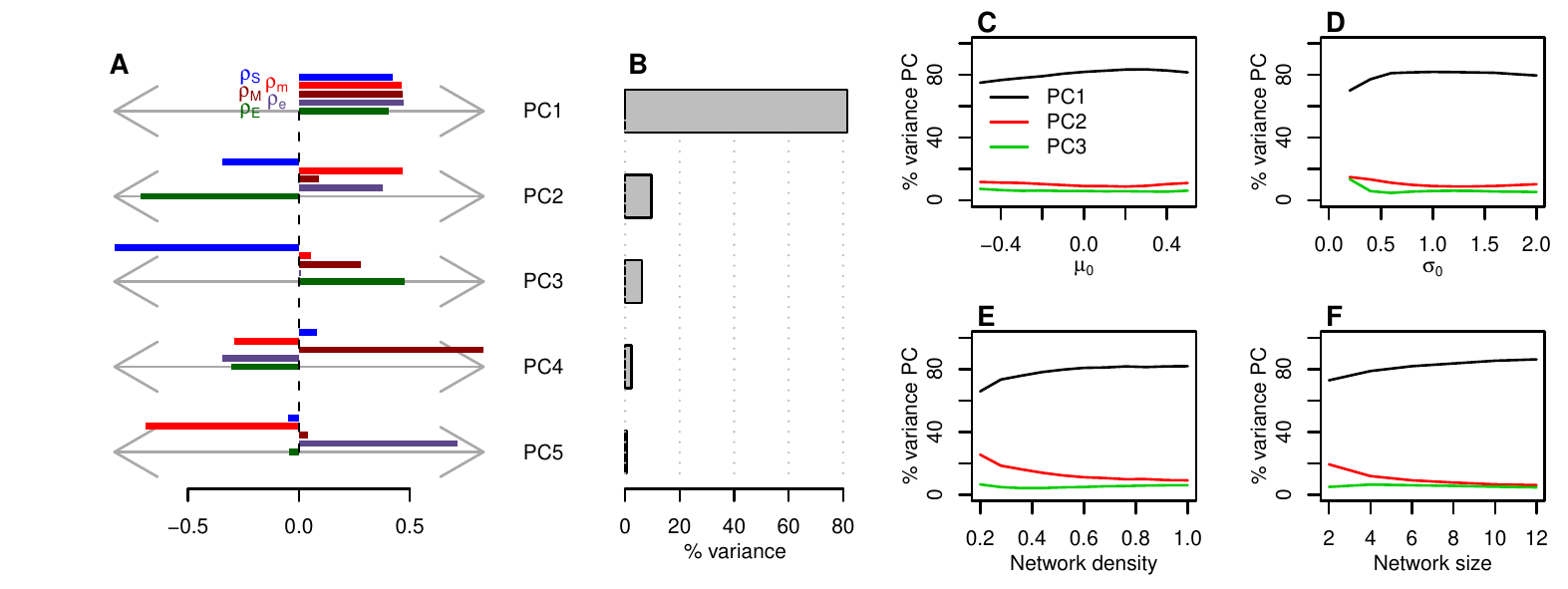}
\caption{\label{fig:pca} A: Summary of the principal component analysis on the five robustness indicators over 10,000 random 6-gene networks ($\mu_0=0, \sigma_0=1$), indicating the position of the five robustness components on all five (normalized) Principal Components (PC); $\stability$: Stability, $\earlyenv$: Early environmental, $\lateenv$: Late environmental, $\earlymut$: Early genetic, $\latemut$: Late genetic. B: relative contribution of the five PCs to the total variance. C: Influence of the average regulation strength ($\mu_0$) on the \% of the total variance explained by the first PC (negative values feature inhibitory networks, positive values activating networks). D: Influence of the standard deviation of the regulation strength ($\sigma_0$). E: Influence of the network density. F: Influence of the network size. }
\end{adjustwidth}
\end{figure}

Correlations were calculated between all five robustness components over 10,000 random networks (\SupMat \ref{supp:fullcorr}). All robustness components were positively correlated, correlations ranged from about 0.62 (late genetic vs.\ early environmental) to above 0.97 (late environmental vs.\ late genetic). A Principal Component Analysis (Figure~\ref{fig:pca}A and B) confirms that robustness components were partially correlated. The first PC (82\% of the total variance) corresponds to the general robustness of the network, and involves all robustness indexes. The remaining variance is explained by orthogonal vectors separating all other robustness components. At least 4 out of 5 PCs, explaining 10\% to 2\% of the total variance, did not vanish when increasing the sample size (\SupMat \ref{supp:PC}). The part of the variance in robustness explained by the first PC is robust to the network properties, as it remains around 80\%  when the mean and the variance in the regulation strengths, the network density, and the network size vary (Figure~\ref{fig:pca}C, D, E, and F).  \\

\subsection*{Two-gene networks}

\begin{figure}[t]
%~ \captionsetup{justification=centering}
\begin{adjustwidth}{-1.5in}{0in}
\includegraphics{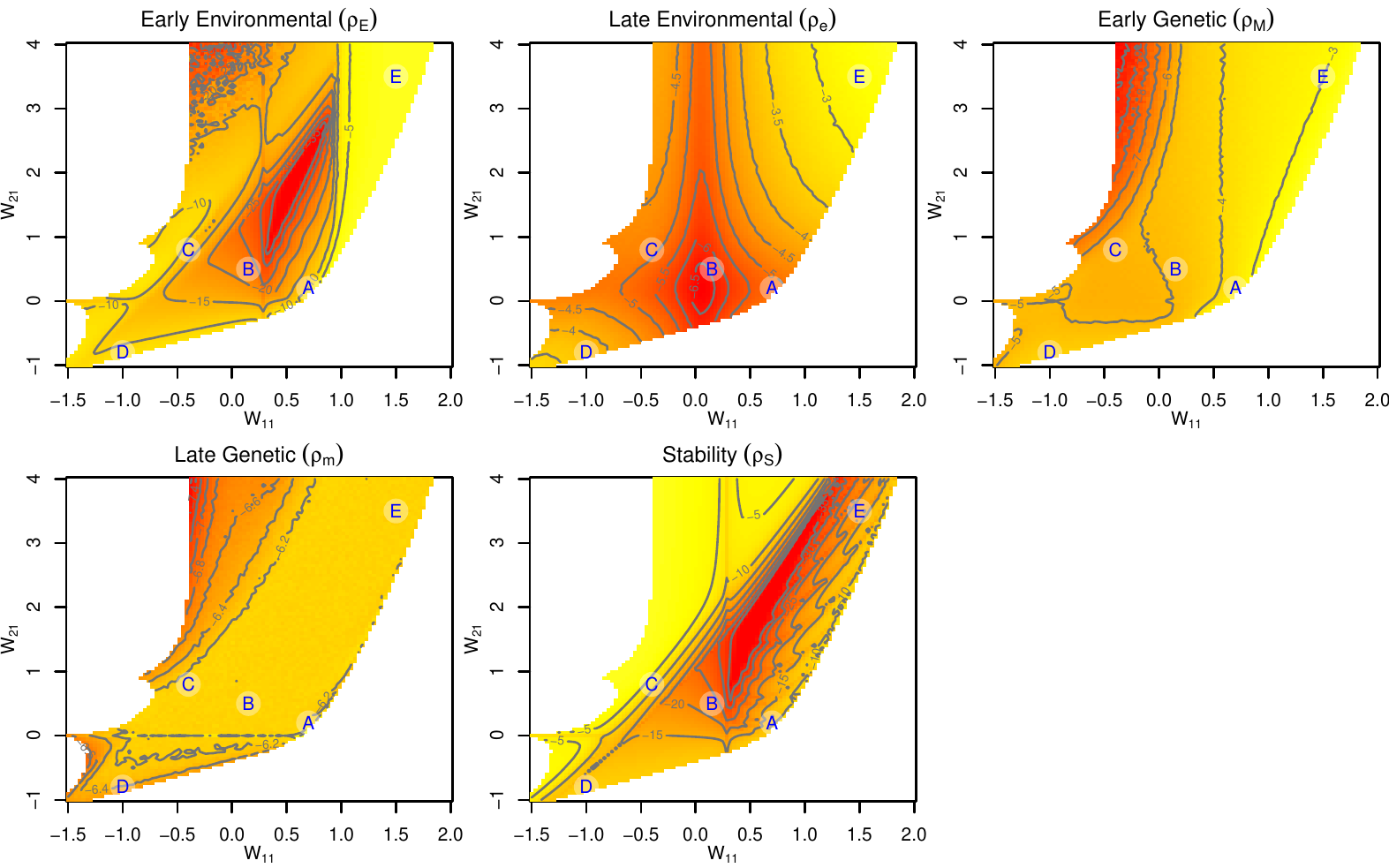}
\caption{\label{fig:imgpanels} Robustness indicators ($\earlyenv$, $\lateenv$, $\earlymut$, $\latemut$, and $\stability$) estimated for an exhaustive continuum of two-gene networks with an arbitrary expression equilibrium at $\Pp_\infty = (0.3, 0.6)$. Although two-gene networks have four independent genetic parameters, only two were represented here, the two others being computed to ensure the desired equilibrium. Red stands for the maximum robustness (lowest robustness scores); yellow for minimum robustness (highest scores). For readability, color scales are different across panels. Letters A to E stand for five example networks illustrated in \SupMat \ref{supp:simpanels}.}
\end{adjustwidth}
\end{figure}

In the following, I considered an arbitrary case of a two-gene network which genes are expressed to $\Pp_\infty$ = (0.3, 0.6). Equivalent results could be achieved with a different, arbitrary target. Figure~\ref{fig:imgpanels} illustrates how the robustness components varied in this constrained 2-gene network model (red stands for maximum robustness, i.e.\ minimum scores for $\stability$, $\earlyenv$, $\lateenv$, $\earlymut$, and $\latemut$). All the networks considered here converge to the same gene expression, and can thus be considered as phenotypically equivalent ; the colored space in Figure~\ref{fig:imgpanels} thus represents a connected neutral network in which populations can evolve, and thus change the topology and the robustness of the gene network, while keeping the expression phenotype constant. In the white regions, the equilibrium was not achieved in numerical simulations for at least three different reasons (\SupMat \ref{supp:whyitfails}): (i) fluctuations around the equilibrium were large enough to hit the edges of the (0,1) interval, shifting the mean expression; (ii) the expression dynamics was slow and the network was unable to get close to the equilibrium after 16 time steps; (iii) the equilibrium was not reachable from the default starting point. \\

The different robustness components were correlated, but did not overlap perfectly. In order to assess the variation of the robustness properties, five networks of contrasted robustness, labeled from A to E, were tracked more specifically (Figure~\ref{fig:imgpanels}; the corresponding $\W$ matrices are provided in \SupMat \ref{supp:W}). \SupMat \ref{supp:simpanels} illustrates the effect of various sources of disturbance on each network dynamics. The network denoted as~B was robust to most sources of disturbance, while network~E was sensitive to all components except stability. Network~C was unstable, but remained relatively buffered. Networks~A and~D illustrate intermediate loss-of-robustness behaviors, through different mechanisms (instability for network~D, and weak buffering for network~A).  \\

This 2-gene network analysis thus confirms the results obtained for large random networks: robustness components are only partially correlated. Robustness is not a feature of large and intricate genetic architectures, as it is already present (and multidimensional) in the simplest gene networks.  \\

\subsection*{Evolution and evolvability of robustness}

The evolution of robustness was studied by individual-based simulations, in which all individuals were characterized by their genotype (a 6-gene network) and a set of phenotypes (gene expressions and network robustness). Gene expressions for 3 out of 6 genes were under stabilizing selection. In addition to stabilizing selection on gene expression (forcing the network to maintain a functional role), robustness indicators were directly selected towards more or less sensitivity. Such direct, artificial selection pressures on robustness are not designed to reflect realistic selection on gene networks, but they might reveal evolutionary limits to the evolution of robustness due to internal constraints. Stabilizing selection on gene expression is expected to generate a slight selection pressure on the robustness, but this effect was apparent only for larger or more frequent mutations (\SupMat \ref{supp:explo}).  \\

\begin{figure}[tp!]
\begin{adjustwidth}{-1.5in}{0in}
%~ \captionsetup{justification=centering}
\includegraphics{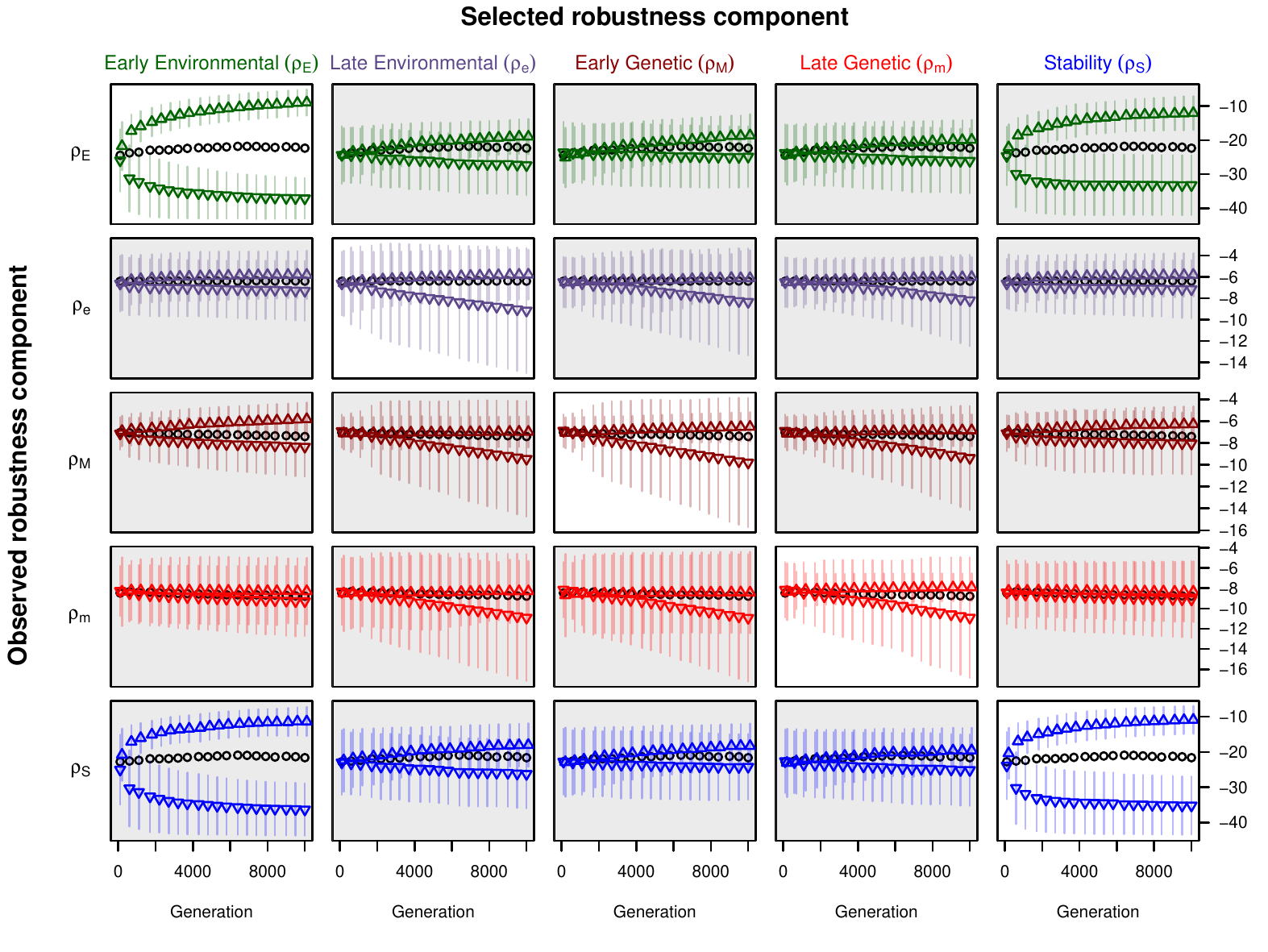}
\caption{\label{fig:evol} The evolution of all five robustness indicators under direct and indirect selection was recorded for 10,000 generations in individual-based simulations. The figures show the average and standard deviation of robustness over 100 replicated simulations. The control simulations (black circles) correspond to stabilizing selection on gene expression only (no direct selection on robustness). Colored symbols correspond to simulations in which robustness indicators were selected up or down (upward or downward triangles), colors correspond to the observed indicator, columns indicate which indicator was selected (diagonal panels: direct selection, off-diagonal panels: indirect selection).}
\end{adjustwidth}
\end{figure}

Direct selection on all robustness components lead to a response, showing that robustness is evolvable (diagonal panels in Figure~\ref{fig:evol}). Yet, the evolutionary potential differed substantially among robustness indicators, as indicated by the differences in the Y-scales. Robustness indicators being all homogeneous to a sum of squared difference in gene expression (i.e., the variance in gene expression induced by various disturbances), they could be compared directly. The most evolvable robustness components were early environmental disturbances ($\earlyenv$) and stability ($\stability$), which can differ by up to 25 log units (11 orders of magnitude) after 10,000 generations of bidirectional selection. In contrast, robustness to late environmental noise $\lateenv$ and genetic changes ($\earlymut$ and $\latemut$) only differed by 3 to 4 log units (i.e.\ a factor 10 to 100). For these three robustness components ($\lateenv$, $\earlymut$, and $\latemut$), the response was clearly asymmetric (the response towards more robustness was slower). Although the average response supports a clear evolutionary trend, response to selection was variable across simulation replicates, as distributions of up and down responses generally overlap. The selection response was still ongoing after 10,000 generations.  \\

\begin{figure}[thp!]
\begin{adjustwidth}{-1in}{0in}
%~ \captionsetup{justification=centering}
\includegraphics{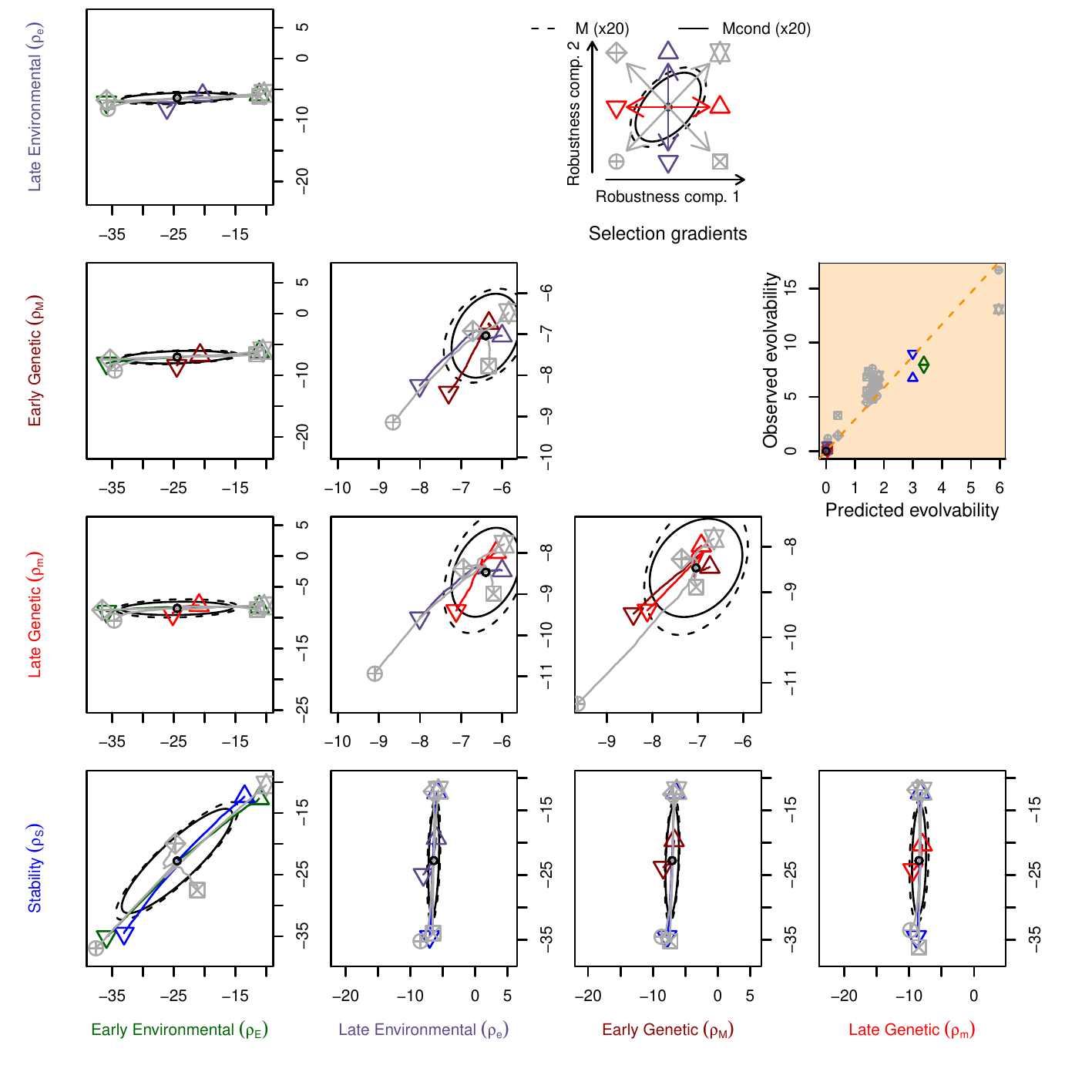}
\caption{\label{fig:evolvability} Trajectories of the bivariate response to selection over 5000 generations (average over 100 simulation replicates) for all combinations of robustness indicators. Each panel displays the selection response in eight directions, as illustrated in the legend (four univariate --- colored arrows --- and four bivariate --- gray arrows --- gradients of selection, same color code as in Figure~\ref{fig:evol}). Mutational and conditional mutational matrices, estimated from the initial genotypes, are illustrated as ellipses in each panel (95\% ellipses assuming a multivariate Gaussian mutational distribution). For conditional $\bm{\mathrm M}_c$ matrices, the constraining traits were the three gene which expression was under stabilizing selection. X and Y axes were adjusted so that their scale matches for each trait comparison (correlational ellipses were not distorted). The colored inset illustrates the proportionality between the predicted mutational evolvability (calculated from $\bm{\mathrm M}_c$) and the observed evolvability in the direction of selection after 1000 generations (same color/symbol code as in the rest of the figure, hyphenated line: linear regression with no intercept). }
\end{adjustwidth}
\end{figure}

Selection on robustness components also lead to an indirect response of all other components, which confirms a general genetic correlation. The magnitude of the correlated response (from 10\% to 100\% of the direct response) depended on the correlation across robustness components. Simulations were run to test the long-term effect of synergistic and antagonistic selection on all pairs of robustness indicators (Figure~\ref{fig:evolvability}), and selection responses were compared to the mutational evolvabilities computed at the beginning of the simulations. There was a convincing proportional relationship between predicted and observed evolvabilities on all directions of selection. Selection response was fast in directions that were mutationally evolvable, and slow in directions that were not evolvable. Yet, in spite of the variation of evolvability across directions in the multivariate robustness space, evolution was always possible, even if reduced proportionally to the mutational variance, confirming the absence of absolute constraints.  \\

\begin{figure}[thp!]
%~ \captionsetup{justification=centering}
\includegraphics{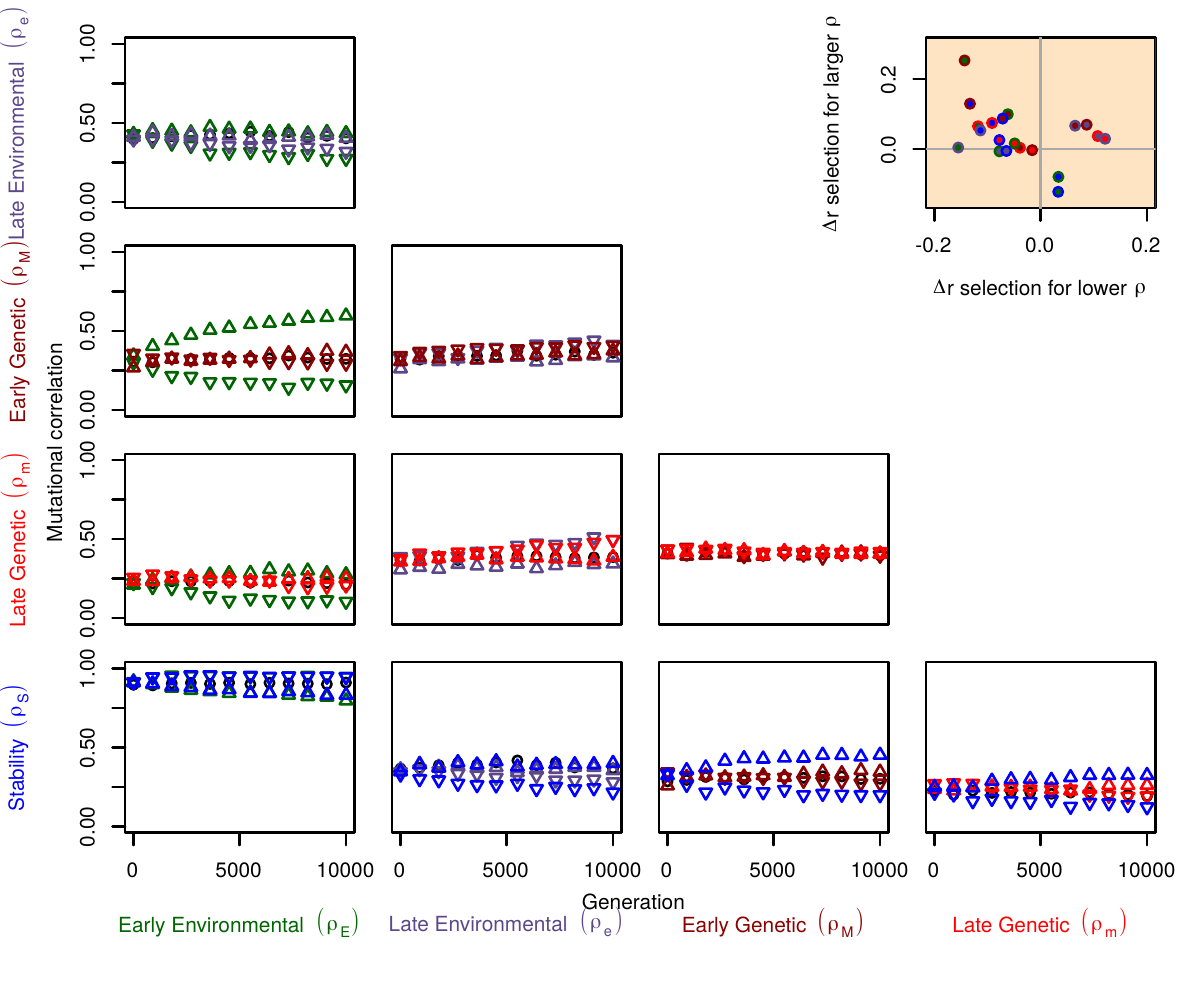}
\caption{\label{fig:evolcor} Evolution of the average mutational correlation under univariate selection on robustness components. Panels, symbols, and colors are the same as in Figure~\ref{fig:evolvability}. Correlations were estimated for each $\M$ matrix and averaged over 100 simulation replicates. The color insets shows the (lack of) consistency between the evolution of correlations when selecting for higher (x axis) and lower (y axis) robustness. $\Delta r$ stands for the difference between correlation at generation 10,000 and at generation 1. In the inset, the color of the symbol filling corresponds to the selected robustness component, the color of the symbol border to the correlated robutsness component.}
\end{figure}

The proportionality between realized and predicted evolvabilities tends to fade out for long-term selection responses (\SupMat \ref{supp:r2evolv}), which can be due to the evolution of mutational constraints (the $\M$ matrix evolves compared to the initial network). This was confirmed by tracking the evolution of mutational correlations across robustness traits through time (Figure~\ref{fig:evolcor}). Average correlations did not evolve substantially in control simulations, but direct selection on robustness components did trigger systematic change in some (but not all) mutational correlations. For instance, the correlation between $\earlymut$ and $\latemut$ does not seem to be evolvable, while the correlation between $\earlymut$ and $\earlyenv$ changed from $\simeq 0.3$ to about $0.6$ or $0.15$ depending on the selection regime. All correlations remained positive. The evolution of correlations was partially driven by the direction of selection (more or less robustness). Within each specific pair of robustness components, the evolution of correlation was rather consistent: for instance, selecting to decrease $\earlyenv$ or $\earlymut$ (i.e.\ making the network more robust) always decreased the correlation between $\earlyenv$ and $\earlymut$. Yet, there was no general pattern associating the evolution of robustness and the evolution of correlation; depending on the robustness component, selecting for more or less robust networks may increase or decrease the correlations (colored inset in Figure~\ref{fig:evolcor}). There was no effect of joint selection; selecting together two robustness components did not make them more (or less) correlated (\SupMat\ref{supp:cordirsel}). \\

\section*{Discussion}

Whether or not various robustness components of genetic architectures are independent is central to understand why organisms are robust or sensitive to genetic or environmental disturbances. Independent genetic bases of robustness components would call for independent evolutionary histories, while a pleiotropic genetic architecture could explain the evolution of nonadaptive robustness components as a result of indirect selection. The analysis of the genetic correlations between five robustness components, based on a simple gene network model, results in a balanced answer: robustness components are largely correlated, but pleiotropy is not an absolute constraint, and pairs of robustness components evolved in divergent directions under direct, artificial bivariate selection. Such a quantitative answer to the so-called 'congruence' hypothesis \parencite{dHW+03} would explain both how unselected robustness components could be partly driven by indirect selection and why various robustness-related features seem to have their own evolutionary history.  \\

\subsection*{Model limits}

Gene regulation networks are popular candidates when attempting to model complex biological processes: they are at least partly built on solid and realistic principles (transcription factors can enhance or repress the expression of other genes), gene regulation plays a crucial role in most biological, physiological, and developmental mechanisms, and even modest size regulation networks display a wide diversity of behavior, including homeostasis (stable equilibrium of gene expressions) \parencite{Ste99}, cyclic dynamics \parencite{LG03,ARB+10}, or amplification of a weak signal \parencite{HB08}. Conveniently, the phenotypic level considered as the output of a gene network (the expression level of all network genes) can be assimilated to a partial transcriptome, which opens the possibility for confrontation with empirical data.  \\

The gene network model proposed by \textcite{Wag94} is particularly popular in evolutionary biology to model gene network evolution due to its computational simplicity and efficiency, combined with a direct biological interpretation (each line of the regulation matrix is the set of transcription factor fixation sites in the promoter of a gene) (see \textcite{SH13,FP15} for review and alternative models). In practice, multiple variants based on this original model have been derived, either to address specific questions, or to correct for unrealistic features. Here, I used a quantitative version of the model, in which gene expressions were scaled between 0 (no expression) and 1 (maximum expression), which was first proposed in \textcite{Wag94}, although later work have often preferred binary networks (in which genes can be on/off, e.g. \textcite{Wag96,CMW07}), and a gene expression scaling between -1 and 1. Unlike in \textcite{Wag96, SB02}, mutations had cumulative effects (the value of the mutant allele was drawn in a Gaussian centered around the value of the parental allele), which allows for gradual evolution. Finally, the sigmoid response function was made asymmetrical by introducing a constitutive expression parameter (as in e.g. \textcite{RL16}) in order to avoid the unrealistically high expression of unregulated genes (half the maximum expression) from the default setting. This constitutive expression was not evolvable in the model, but simulations (\SupMat \ref{supp:explo}) show that two robustness components ($\earlyenv$ and $\stability$) were very sensitive to this parameter (larger constitutive expression was associated with more robust networks). It is thus not unlikely that real systems may evolve towards more robustness by increasing the constitutive expression of key genes, as already suggested (for different reasons) by \textcite{DW15}.  \\

Discrete time and simple matrix algebra made it possible to run evolutionary individual-based computer simulations, in which the network output needs to be calculated for thousands of individuals and thousands of generations. Using more realistic models based on continuous time and differential equations, non-linear regulation effects, and independent degradation and transcription rates would make the simulations less practical, with little benefit in terms of explanatory power. Computational constraints also limit the network size to a few dozen genes, which was not enough to generate realistic levels of sparsity --- simulated gene networks were too dense to be realistic. Decreasing network density and smaller network sizes made robustness components slightly less correlated (Figure~\ref{fig:pca}E and~F), suggesting that the integration of robustness components increases with network complexity (size and number of connections). The simulated phenotypic target (maintaining a constant set of gene expressions) were also extremely simple compared to what gene networks are theoretically able to do (e.g. converging to different equilibria in different cell types, or controlling a complex dynamic of gene expression during the development). However, the results are robust to most simulation parameters (\SupMat \ref{supp:explo}), suggesting that they reflect general properties of the underlying genetic architecture. In particular, the network size $n$ and the number of selected genes $n^\prime$ do not alter drastically robustness components, showing that small regulatory motifs are not qualitatively different from large gene networks in terms of robustness.  \\

In spite of the simplicity of the network model, it appeared that connecting network features (for instance, the strength of a specific regulation) and robustness was not trivial, even in very small networks. For instance, in the $n=2$ gene-network analysis, most robustness components were complex functions of all four regulation strengths. Throughout this work, robustness was thus treated as an emergent property of the underlying network, which cannot be easily deduced from a reductionist approach. Yet, it is possible to interpret the correlation patterns in terms of network dynamics. Two of the most correlated components are the robustness to early environmental variation $\earlyenv$ and network stability $\stability$, which both measure the ability of the network to converge to a given gene expression equilibrium. Conversely, the correlation between late mutational $\latemut$ and environmental $\lateenv$ robustnesses can be attributed to the consequences of such disturbances over a single time step: for a single target gene, decreasing the concentration of a transcription factor and decreasing the sensitivity of the promoter to the same transcription factor have very similar immediate consequences on gene expression. Yet, even if these measurements happen to be correlated by construction, their partial evolutionary independence highlights their potential for independent evolvability in real gene network architectures, which are substantially more complex and subtle than our gene network model.  \\

In the simulations, selection on robustness components was direct and constant both in up and down directions (i.e.\ towards more or less robust genetic architectures). This setting was not expected to reflect realistic evolutionary pressures on robustness, which might be more complex, overlapping, and asymmetric. Stabilizing selection, for instance, selects both directly for robustness to environment, and indirectly for robustness to mutations \parencite{WBB97}; selection for stability also promotes indirectly robustness to mutations \parencite{SB02}. Conversely, selecting for lower robustness through the phenotype may be difficult or even impossible: fluctuating selection does not promote decanalized genetic architectures \parencite{LAH13}, and selection for environmental sensitivity is limited by the inaccuracy of the perception of the envrionmental signal \parencite{RWS+10}. Simulation results thus illustrate how robustness components may evolve independently when individually selected; whether or not there exists realistic conditions for such selection pressures is a different --- and more complicated --- issue.  \\

\subsection*{Measuring robustness}

There are potentially many ways to measure the robustness of a phenotypic trait. Here, five indicators were proposed to capture various (and potentially independent) aspects of what is generally defined as robustness. The sensitivity to inherited mutations ($\earlymut$) is probably the most popular one, as it is central to the discussion around the evolution of canalization \parencite{Wad59, Wag96, Far15}. The sensitivity to environmental perturbations is also unavoidable, although its implementation in a gene network model is less straightforward. Here, it was calculated as both the sensitivity of the network to disturbance in the initial expression state ($\earlyenv$), which measures the size of the basin of attraction of the optimal expression pattern, and as the strength of the stability of the equilibrium when disturbed ($\lateenv$). These two measurements can be interpreted as developmental robustness and physiological homeostasis, respectively, as they quantify the response of the network to disturbances in the expression levels at different time scales. The robustness to mutations occurring after the network convergence ($\latemut$) was considered because it sets up an alternative to the genetic vs.\ environmental congruence hypothesis: in long-lived organisms, non-heritable (somatic) mutations participate to the ageing process \parencite{KLH12}, ageing being to some extent under direct selection. Thus, the robustness to somatic mutations could also drive indirectly the evolution of genetic canalization. Although not strictly a robustness component, the gene network stability ($\stability$, amplitude of the fluctuations of gene expressions) was also considered because it has been proven to drive an indirect response of genetic canalization, based on very similar model \parencite{SB02}. Its correlation with other robustness indicators confirms the tight link between robustness and stability in gene networks. \\

These indicators were chosen based on the possibility to measure them in numerical simulations. Although the empirical assessment of the correlation between robustness components would be way more convincing than a theoretical study, defining similar measurements from experimental datasets can be challenging. For instance, $\earlymut$ and $\earlyenv$ could, at least in theory, be estimated as the variance in gene expression across genetic backgrounds or across environmental conditions, respectively. Measuring $\latemut$ environmentally is more complicated, as it would likely be confounded with other ageing mechanisms. In contrast, the empirical distinction between e.g.\ $\lateenv$ and $\stability$ relies on discriminating internal vs.\ external sources of noise, and might be in practice impossible. In all cases, gene expression data are generally quite noisy and their analysis necessitates heavy corrections to prevent multiple testing issues. Studying empirically the robustness and evolvability of molecular and morphological traits has long been considered as a challenging task, but methodological and technological progress has recently brought new concrete perspectives \parencite{PW19}.  \\

Some popular measurements of developmental robustness were not considered here for technical reasons. For instance, fluctuating asymmetry (the variance between the same phenotypic trait measured in the right and the left body parts of symmetric organisms) is a convenient measurement of microenvironmental effects on the development \parencite{DD01,LK05}, but it has no equivalent at the level of gene expression in a regulation network. The deterministic sensitivity to a directional environmental gradient could also be used to measure phenotypic plasticity, which is central to the question of phenotypic robustness. Yet, there are several ways to model phenotypic plasticity in a gene network \parencite{Mas04,BTL21}, and it requires a specific selection setup (different expression optima as a function of the environment). Because of this additional complexity, adaptive phenotypic plasticity was excluded from the focus of this work, although the evolution of plasticity of gene expression remains an intriguing and fundamental question. In particular, phenotypic plasticity (i.e. an adaptive lack of robustness to some environmental signal) may itself be canalized to genetic or other environmental disturbances \parencite{SK94}; considering reaction norms (a measurement of plasticity) as quantitative traits thus opens challenging questions about the adaptive evolution of the canalization of robustness traits. \\

\section*{Data accessibility}
Scripts to reproduce simulations and figures are available online (\url{https://doi.org/10.5281/zenodo.6393075}), \textcite{GitHubRepos}. 

\section*{Acknowledgements}
Version 5 of this preprint has been peer-reviewed and recommended by Peer Community In Evolutionary Biology (\url{https://doi.org/10.24072/pci.evolbiol.100138}). I thank the recommender and the three reviewers for their constructing comments which have substantially improved the manuscript. Many thanks to Laurent Loison for insightful discussions. Simulations were partly performed on the Core Cluster of the Institut Français de Bioinformatique (IFB) (ANR-11-INBS-0013) .

\section*{Funding}
No specific funding beyond the CNRS basic support to the author. 

\section*{Conflict of interest disclosure}
The author of this preprint declare that they have no financial conflict of interest related to the content of this article. 
%%%%%%%%%%%%% THE TEXT ENDS ABOVE THIS LINE %%%%%%%%%%%%%%%%%%%%%%%

\printbibliography[notcategory=ignore]

\section*{Appendix}
  \setcounter{section}{0}
  \renewcommand{\thesection}{Appendix~\arabic{section}}

  \section{}
    \label{supp:sensitmag}
    \subsection*{Sensitivity of the robustness measurements to the magnitude of the disturbance}
	\begin{center}
	\includegraphics{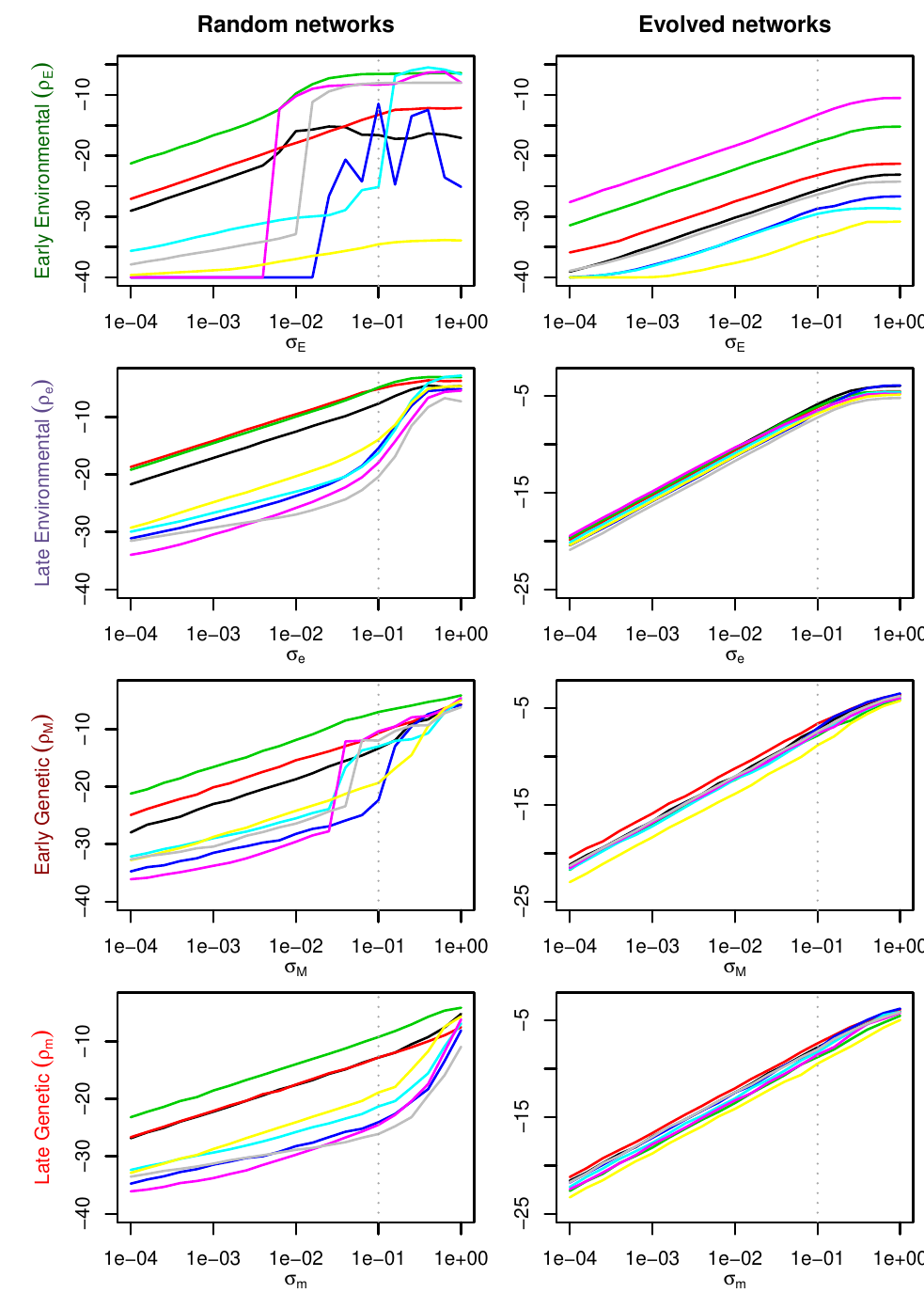}
	\end{center}
	\clearpage
	{Four out of five robustness indicators ($\earlyenv$, $\lateenv$, $\earlymut$, $\latemut$) depend on the magnitude of the disturbance ($\sigma_E$, $\sigma_e$, $\sigma_M$, and $\sigma_m$, respectively). The figure displays the influence of the size of the disturbance on the robustness measurement (left: 10 random networks, right: 10 evolved networks). Vertical dotted lines stand for the values used in the simulations. Robustness scores are not completely consistent for random networks, as some of them can be differentially robust to large or small disturbances. The consistency is better in evolved networks (the rank of different genotypes in terms of robustness rarely depends on the size of the disturbance).  }

  \clearpage
  \section{}
    \label{supp:fullcorr}
    \subsection*{Correlations among robustness indexes among random networks}
	\begin{adjustwidth}{-1in}{0in}
	\begin{flushright}
	\includegraphics{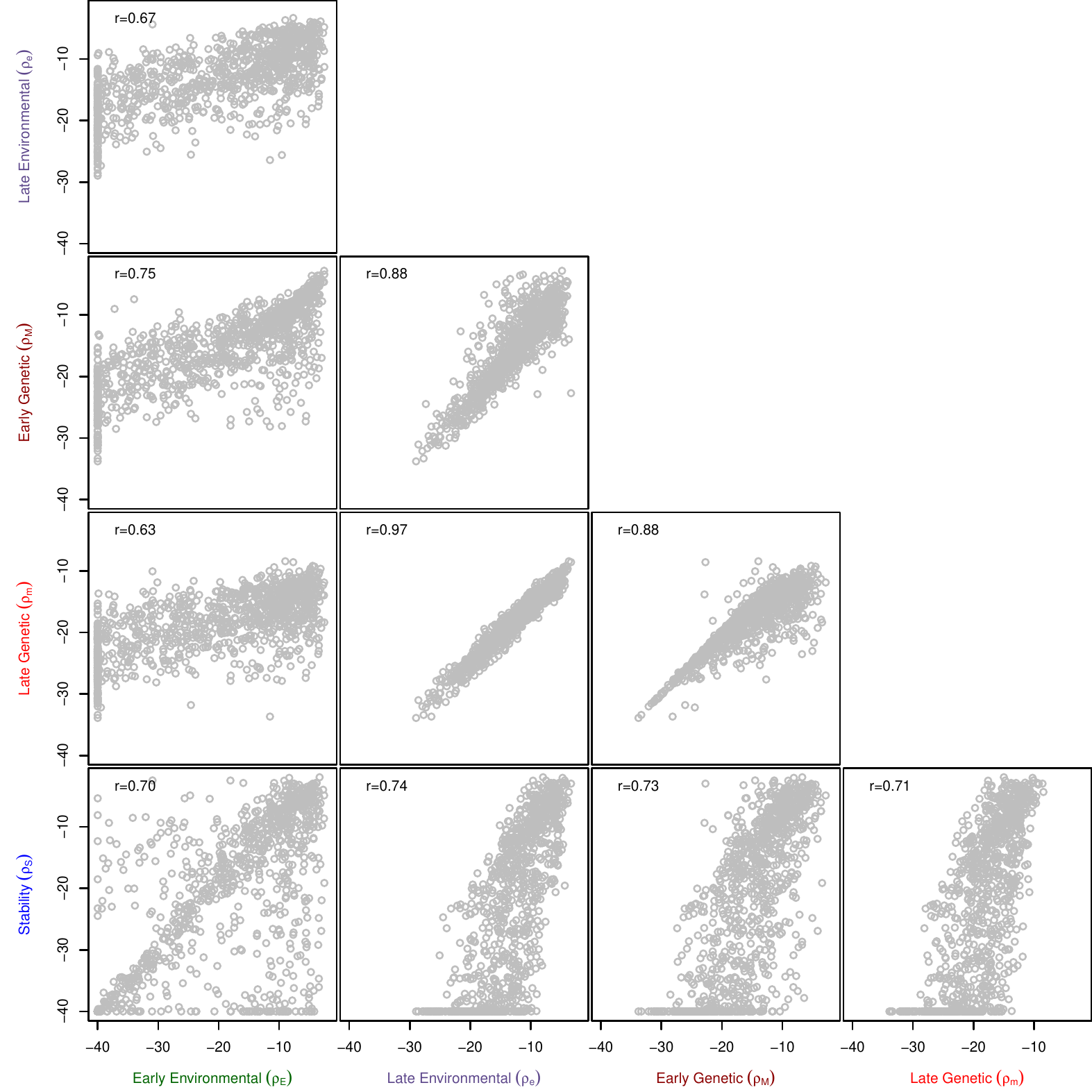}
	\end{flushright}
	\end{adjustwidth}
	Correlations between all five robustness components among 10,000 random 6-gene networks ($\mu_0=0, \sigma_0=1$). 

  \clearpage
  \section{}
    \label{supp:PC}
    \subsection*{Sampling effects on Principal Components}
    
	\begin{center}
	\includegraphics{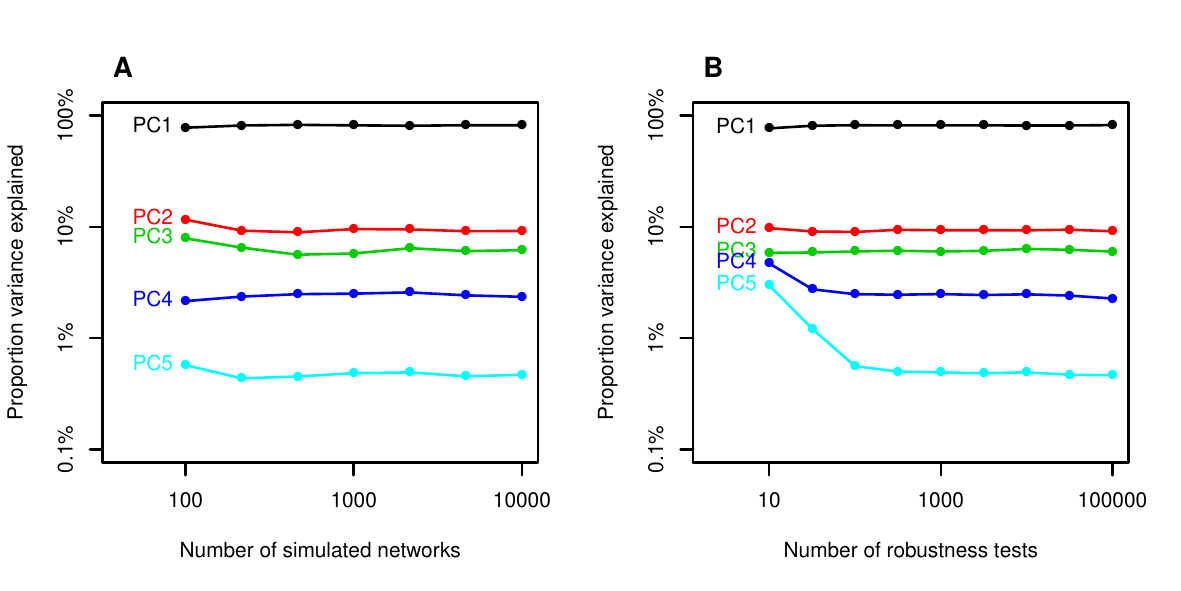}
	\end{center}    

	{Influence of the sampling effect (number of networks and number of replicates $R$ to estimate robustness) on the relative
	weight of the principal components. All PCs except the last one are robust to sampling. }

  \clearpage
  \section{}
    \label{supp:whyitfails}
    \subsection*{Reasons for not reaching the desired equilibrium}
	\begin{center}
	\includegraphics{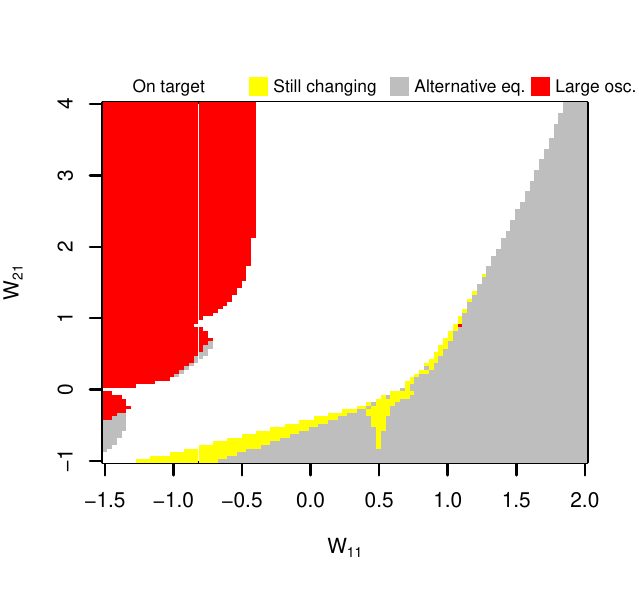}
	\end{center}
	
	{Although equation~\ref{eq:predef} guarantees that an equilibrium exists at the target phenotypic expression, the equilibrium might not be reachable in practice when simulating the gene network dynamics. The colored area in the figure corresponds to networks that failed to produced the target phenotype, each color representing a distinct reason; Yellow: network dynamics was slow and the final gene expression has not been reached yet after 16 time steps; Gray: an alternative equilibrium was reached (most of the time implying that one or both genes are either completely silenced to fully expressed). Red: The network steady state featured oscillations that were so large that they hit the maximum or minimum expression, shifting the average expression away from the target expression. }

  \clearpage
  \section{}
    \label{supp:W}
    \subsection*{Two-gene example networks}

	\begin{center}
	\begin{tabular}{rrrrr}
	  \hline
	 & $W_{11}$ & $W_{21}$ & $W_{12}$ & $W_{22}$ \\ 
	  \hline
	  A & 0.70 & 0.20 & -0.21 & 0.38 \\ 
	  B & -0.30 & 0.30 & 0.29 & 0.33 \\ 
	  C & -0.40 & 0.80 & 0.34 & 0.08 \\ 
	  D & -1.00 & -0.80 & 0.64 & 0.88 \\ 
	  E & 1.50 & 3.50 & -0.61 & -1.27 \\ 
	   \hline
	\end{tabular}
	\end{center}
	{The five two-gene networks detailed in Figure~\ref{fig:imgpanels} and \SupMat \ref{supp:simpanels}.}

  \clearpage
  \section{}
    \label{supp:simpanels}
    \subsection*{Illustration of the robustness scores}

	\begin{adjustwidth}{-1in}{0in}
	\includegraphics{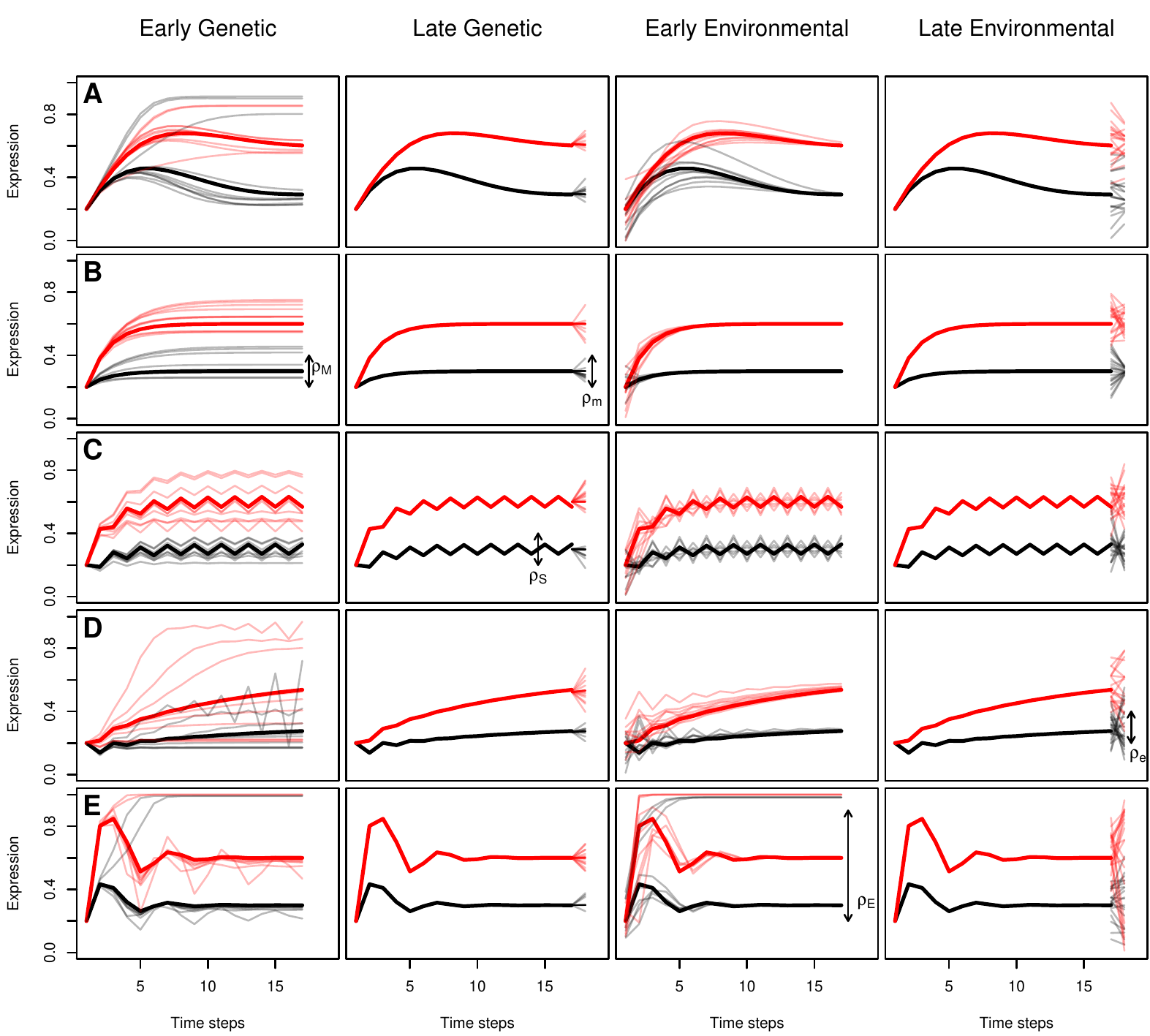}
	\end{adjustwidth}
	\clearpage
	{The figure displays a subset of the replicated tests for four robustness indexes. Rows A to E correspond to the five networks described in \SupMat \ref{supp:W}. Four (out of five) robustness measurements rely on a resampling procedure (corresponding to the four columns of the figure). In each panel, the default (undisturbed) network kinetics is displayed as plain lines (black for gene~1, red for gene~2), while 10 disturbed networks are indicated as pale lines. By construction, all networks have an equilibrium at (0.3, 0.6). The network stability can be assessed from the amplitude of the cycles in the undisturbed kinetics (thick lines), and does not rely on a stochastic algorithm. The network robustness to genetic disturbance was estimated by mutating the gene network before the first time step (early genetic mutation, first column) or before the last time step (late genetic mutation, second column). Environmental robustness was estimated by disturbing the gene expression, without changing the genotype, before the first time step (early environmental, third column) and before the last time step (late environmental, fourth column). }

  \clearpage
  \section{}
    \label{supp:explo}
    \subsection*{Exploration of the parameter set}
	\begin{adjustwidth}{-2.0in}{0.5in}
	\includegraphics{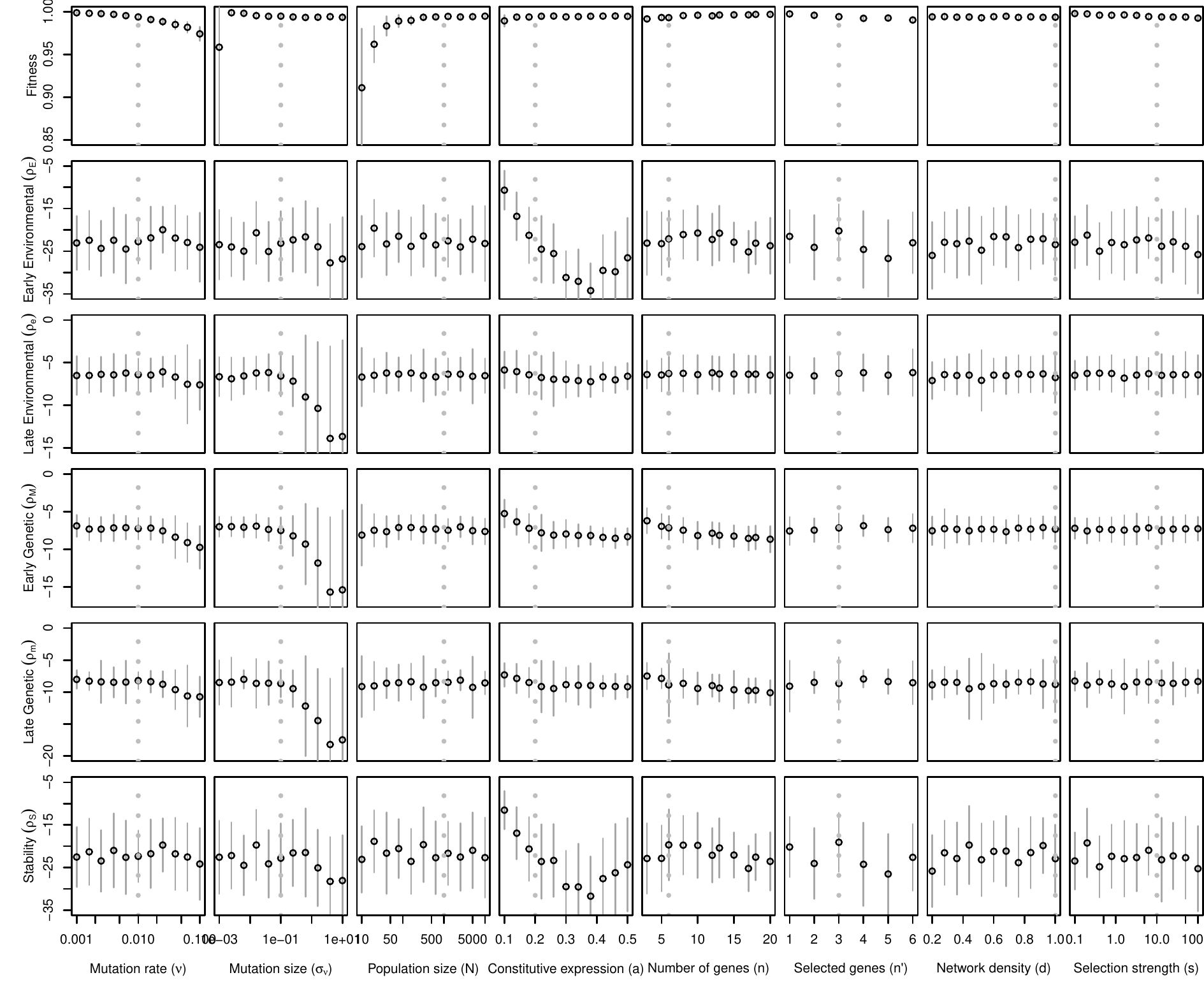} 
	\end{adjustwidth}
	\clearpage
	{Influence of simulation parameters (mutation rate $\nu$, mutation size $\sigma_\nu$, population size $N$, constitutive expression $a$, total number of genes $n$, number of selected genes $n^\prime$, network density $d$, and strenght of selection $s$) on fitness and robustness indexes after 5000 generations (default settings except for the target parameter). The figure reports the mean $\pm$ standard deviation across 20 replicated simulations. Vertical dotted lines stand for the default parameter values. }

  \clearpage
  \section{}
    \label{supp:r2evolv}
    \subsection*{Accuracy of the prediction vs.\ simulation time}
	\begin{center}
	\includegraphics{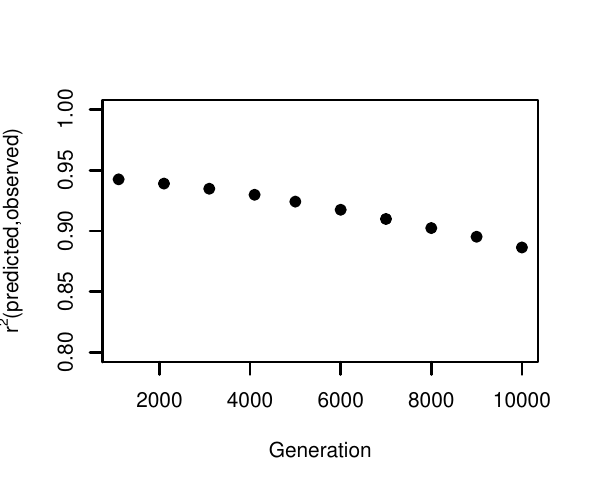} 
	\end{center}
	
	{Effect of the number of generations on the proportionality relationship between predicted and observed evolvabilities of robustness components. The figure displays the $r^2$ of a linear regression (without intercept) between the predicted evolvability from the conditional $\M_c$ matrix measured at the first generation and the observed evolvability in the direction of selection for all replicated simulations. The regression at generation 1,000 is illustrated in the colored inset in Figure~\ref{fig:evolvability}. }

  \clearpage
  \section{}
    \label{supp:cordirsel}
    \subsection*{Evolution of correlations}
    
	\begin{center}
	\includegraphics{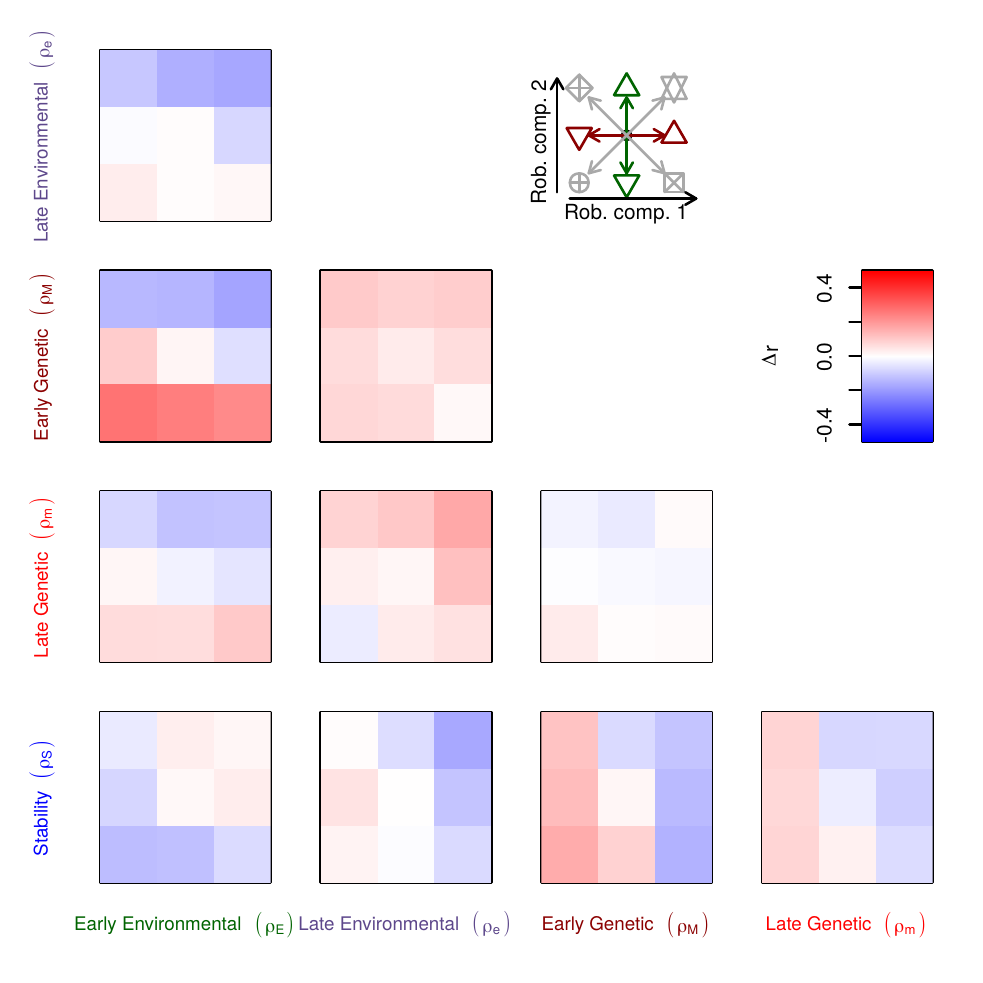} 
	\end{center}
	
	{Evolution of the mutational correlation among robustness components after 10,000 generations of evolution ($\Delta r = r_{10,0000} - r_0$), averaged over 100 simulation replicates. For each pair of robustness components, nine selection gradients were simulated (including control simulations without selection on robustness, central slot). }

\end{document}